\documentclass[twocolumn,aps,showpacs,floatfix,prc]{revtex4}
\usepackage[dvips]{epsfig}

\begin{document}

\title{Search for long lived heaviest nuclei beyond the valley of stability  }

\author{P. Roy Chowdhury$^1$\thanks{E-mail:~partha.roychowdhury@saha.ac.in}, 
C. Samanta$^{1,2}$\thanks{E-mail:~chhanda.samanta@saha.ac.in}, 
D.N. Basu$^3$\thanks{E-mail:~dnb@veccal.ernet.in}}
\address{ $^1$ Saha Institute of Nuclear Physics, 1/AF Bidhan Nagar, Kolkata 700 064, India}
\address{ $^2$ Physics Department, Virginia Commonwealth University, Richmond, VA 23284-2000, U.S.A.}
\address{ $^3$ Variable  Energy  Cyclotron  Centre, 1/AF Bidhan Nagar, Kolkata 700 064, India}

\date{\today }

\begin{abstract}

The existence of long lived superheavy nuclei (SHN) is controlled mainly by spontaneous fission and $\alpha$-decay processes. According to microscopic nuclear theory, spherical shell effects at Z=114, 120, 126 and N=184 provide the extra stability to such SHN to have long enough lifetime to be observed. To investigate whether the so-called ``stability island" could really exist around the above Z, N values, the $\alpha$-decay half lives along with the spontaneous fission and $\beta$-decay half lives of such nuclei are studied. The $\alpha$-decay half lives of SHN with Z=102-120 are calculated in a quantum tunneling model with DDM3Y effective nuclear interaction using $Q_\alpha$ values from three different mass formulae prescribed by Koura, Uno, Tachibana, Yamada (KUTY), Myers, Swiatecki (MS) and Muntian, Hofmann, Patyk, Sobiczewski (MMM). Calculation of spontaneous fission (SF) half lives for the same SHN are carried out using a phenomenological formula and compared with SF half lives predicted by Smolanczuk {\it et~al}. Possible source of discrepancy between the calculated $\alpha$-decay half lives of some nuclei and the experimental data of GSI, JINR-FLNR, RIKEN are discussed. In the region of Z=106-108 with N$\sim$ 160-164, the $\beta$-stable SHN $^{268}_{106}Sg_{162}$ is predicted to have highest $\alpha$-decay half life ($T_\alpha \sim 3.2hrs$) using $Q_\alpha$ value from MMM. Interestingly, it is much greater than the recently measured $T_\alpha$ ($\sim 22s$) of deformed doubly magic $^{270}_{108}Hs_{162}$ nucleus. A few fission-survived long-lived SHN which are either $\beta$-stable or having large $\beta$-decay half lives are predicted to exist near $^{294}110_{184}$, $^{293}110_{183}$, $^{296}112_{184}$ and $^{298}114_{184}$. These nuclei might decay predominantly through $\alpha$-particle emission.
     
\end{abstract}
\maketitle


\section{Introduction}
            The theoretical studies of properties of heaviest nuclei during the past few decades have drawn considerable attention of experimentalists to investigate the existence of superheavy nuclei beyond the valley of stability. Since the macroscopic \cite{my66,kra79} description on the basis of liquid drop model (LDM) does not take the shell effect into account, it fails to explain the variation of fission barrier height of heavy nuclei with the increase of the fissility parameter ($\sim Z^2/A$). However, according to modern nuclear theory, hindrance to the fissioning of heavy nuclei would be enhanced due to the presence of deformed and spherical shell closures. Different semi-microscopic approaches e.g. macroscopic-microscopic model (MMM) \cite{pat291,smso95,my96,cha97} and its modification \cite{mu01} include pairing and nuclear shell effects \cite{str68} to reproduce the properties of ground and deformed states of nuclei. Many purely microscopic \cite{cw96,ru97,gr02,zh02,be01} descriptions like Hartree-Fock-Bogoliubov (HFB) model with zero range forces of Skyrme type \cite{sk58,cw99} or finite range forces of Gogny type \cite{go80,go84} and relativistic mean field (RMF) \cite{ri96,la96} theory predict the possible deformed and spherical neutron shell closures at N=162 and N=184 \cite{mps02} respectively. Since strong influence of nuclear shells \cite{pat89} in the region of superheavy elements might make sufficiently long lived SHN to be observed, the search for heavier elements in the natural samples was started \cite{muz69,fis72,str96,fl83} about thirty years ago.\\

Experimental investigations in finding the SHN around Z=107-118 have been pursued mainly at three different places: Gesellschaft fur Schwerionenforschung (GSI) in Darmstadt (Germany), Joint Institute for Nuclear research (JINR) in Dubna (Russia), and RIKEN, Japan. In the beginning of the 1980's the first observations of the elements with Z= 107-109 were made at GSI \cite{arm03}. In 1994, $\alpha$-decay chains were observed from nucleus $^{269}110$ \cite{ho195} and later on, $\alpha$-decay chains from nuclides $^{271}110$, $^{272}111$, $^{277}112$ \cite{ho295,ho96,ho98}, $^{283}112$ \cite{ho07} were detected at GSI.\\ 

While RIKEN claimed discovery of the  $^{278}113$ SHN \cite{mo041,mo071}, it also reconfirmed the $\alpha$ decay chains from $^{271}110$ \cite{mo043}, $^{272}111$~\cite{mo042} and $^{277}112$ \cite{mo07}. Observations of the $\alpha$ decay chains of nuclei $^{294}118$,$^{290-293}116$,$ ^{288,287}115$,$ ^{286-289}114$, $^{282-284}113$, $^{285,283}112$~\cite{ei07}, $^{278-280}111$,  $^{273,281}110$ \cite{laza96,prl99}, $^{274-276}109$, $^{275}108$, $^{272,270}107$, $^{271}106$ were reported by JINR~\cite{oga06,oga04,og04,oga07}.\\ 

Recently, the $^{270}Hs$ (Z= 108, N = 162) SHN has been produced in the $^{26}Mg + ^{248}Cm$ reaction \cite{dv06}. According to the theoretical calculations \cite{pat291,pat191}, nucleus $^{270}Hs_{162}$ (Z=108) should have the features of ``deformed doubly magic" nucleus. Most of the heaviest nuclei are expected to be deformed due to partial filling of large nuclear shells by outer nucleons. Dvorak $\it et~al.$ measured the energy ($E_\alpha$) of $\alpha$ particle emitted from $^{270}Hs_{162}$ and used the value of $E_\alpha$ to calculate $Q_\alpha$ ($9.02\pm 0.03~MeV$) for the $\alpha$ decay of $^{270}Hs_{162}$. A phenomenological formula \cite{park05} estimated the $\alpha$ decay half-life ($\sim 22~ s$). \\

 Earlier, it was believed \cite{fis72,nil68,nil69,ran74} that traditional spherical superheavy nuclei might form an ``island of stability" centered around $^{298}114_{184}$ separated from the ``peninsula" of known nuclei by a region of deep instability.  Due to both deformed neutron shell and proton shell effects at Z=108 and N=162 the extension of the peninsula of known nuclei might connect the stability island of spherical superheavy nuclei around doubly magic spherical Z=114 proton shell and spherical N=184 neutron shell. Since fission barrier and shell effect play very important role for the existence of long lived superheavy nuclei it is crucial to determine the fission barrier and half life of fissioning nucleus with a good accuracy. It is well known that very small barrier height against fission can break the nucleus into two fragments immediately after it is formed. The $\alpha$ decay of superheavy nuclei \cite{poen86,adndt07,poen106,poen206,jpsj,deli07,zha07} is possible if the shell effect supplies the extra binding energy and increases the barrier height of fission. $\beta$-stable nuclei having relatively longer half life for spontaneous fission than that for $\alpha$ decay indicates that dominant decay mode for such SHN might be $\alpha$-decay. \\

      In our previous works \cite{adndt07,jpsj,prc06,cs07,prc07} we showed the applicability of the microscopic calculation in predicting the $\alpha$ decay half lives of SHN from a direct comparison with the experimental data \cite{oga06,oga04,og04}. However, as a number of SHN were predicted to have relatively large $\alpha$ decay half lives, it is necessary to find out whether those SHN would survive the fission \cite{mo69} and $\beta$-decay.  In such cases those nuclei can be detected in the laboratory through $\alpha$ decay. This work explores the possibility of finding long lived SHN by comparing the calculated $\alpha$ decay half lives ($T_\alpha$) with available theoretical spontaneous fission (SF) half lives \cite{sm95,sm97}, calculated $\beta$-decay half-lives ($T_\beta$) \cite{MNK97} and the experimental data on SF. The $\alpha$ decay half lives of SHN with Z=102-120 are calculated in a quantum tunneling model with DDM3Y effective nuclear interaction using $Q_\alpha$ values from three different mass formulae.\\ 

     A brief outline of the methodology of the present calculation is presented in section~II. Spontaneous fission half lives from both phenomenological \cite{xu05,ren05} and microscopic approach \cite{sm95,sm97} are given in section~III. Finally, in section~IV, results and discussions  
and in section~V, summary and conclusion are presented.

\vspace{-0.52cm}

\section{Formalism}

\vspace{-0.52cm}

    The  $\alpha$ decay half lives are calculated in the frame work of quantum mechanical tunneling of an $\alpha$ particle from a parent nucleus~\cite{prc06}. The details of calculation of the $\alpha$ decay half lives of superheavy nuclei were described in our earlier works \cite{prc06,cs07,prc07}. The required nuclear interaction potentials are calculated by double folding the density distribution functions of the $\alpha$ particle and the daughter nucleus with density dependent M3Y effective interaction. The microscopic $\alpha$-nucleus potential thus obtained, along with the Coulomb interaction potential and the minimum centrifugal barrier required for the spin-parity conservation, form the potential barrier. The spin-parity conservation condition in a decay process is fulfilled if and only if 

\begin{equation}
 {\bf J} =  {\bf J}_1 + {\bf J}_2 + {\bf l},~~~~~~~\pi =  \pi_1.\pi_2.(-1)^l,  
\label{seqn1}
\end{equation}
\noindent
where ${\bf J}$, ${\bf J}_1$ and ${\bf J}_2$ are the spins of the parent, daughter and emitted nuclei respectively, $\pi$,  $\pi_1$ and $\pi_2$ are the parities of the parent, daughter and emitted nuclei respectively, and {\bf l} is the orbital angular momentum carried away in the process. This conservation law, thus, forces a minimum angular momentum to be carried away in the decay process. Consequently, contribution of the angular momentum gives rise to a centrifugal barrier

\begin{equation}
 V_l = \hbar^2 l(l+1) / (2\mu R^2)
\label{seqn2}
\end{equation}
\noindent
where $\mu$ is the reduced mass of the daughter and emitted nuclei system and $R$ is the distance between them.\\ 

The half lives of $\alpha$ disintegration processes are calculated  using the WKB approximation for barrier penetrability. Spherical charge distributions have been used for calculating the Coulomb interaction potentials. Most of the experimental $Q$-values of $\alpha$ decay ($Q_\alpha$) are obtained from experiments done in GSI, Germany and JINR, Dubna. For theoretical Q-values, mass formulae from KUTY \cite{kuty00}, Myers-Swiatecki \cite{ms} and Muntian et al. \cite{mu01,mu103,mu203} are used.\\

The experimental decay $Q$ values ($Q_{ex}$) have been obtained from the measured $\alpha$ particle kinetic energies $E_{\alpha}$ using the following expression 

\begin{equation}
 Q_{ex} = (\frac{A_p}{A_p-4})E_{\alpha} + (65.3 Z_p^{7/5} - 80.0 Z_p^{2/5}) \times 10^{-6} ~\rm MeV
\label{seqn3}
\end{equation}
\noindent
where the first term is the standard recoil correction and the second term is an electron shielding correction in a systematic manner as suggested by Perlman and Rasmussen \cite{Pe57}. $A_p$ and $Z_p$ are mass number and atomic number of parent nucleus respectively.

      The theoretical decay $Q$ values $Q_{th}$ have been obtained from theoretical estimates for the atomic mass excesses~\cite{mu01,kuty00,ms,mu103,mu203} using the following relationship 

\begin{equation}
 Q_{th} = M - ( M_\alpha + M_d) = \Delta M - (\Delta M_\alpha + \Delta M_d)
\label{seqn4}
\end{equation}
\noindent
which if positive allows the decay, where $M$, $M_\alpha$, $M_d$ and $\Delta M$, $\Delta M_\alpha$, $\Delta M_d$ are the atomic masses and the atomic mass excesses of the parent nucleus, the emitted $\alpha$ particle and the residual daughter nucleus, respectively, all expressed in the units of energy. As $Q_\alpha$-value appears inside the exponential integral as well as in denominator of the expression \cite{prc06,cs07} of $\alpha$ decay half lives, the entire calculation is very sensitive to the Q-values. 

\section{Phenomenological and microscopic calculations for spontaneous fission half lives}

Spontaneous fission of heavy nuclei was first observed by Flerov and Petrjak in 1940 \cite{fl40} from $^{238}U$ nucleus. Spontaneous fission and $\alpha$ decay are the main decay modes \cite{dc89,dc96,ba06} of superheavy nuclei. $\beta$-decay could be another possible decay mode for the superheavies lying beyond the $\beta$-stability line of the nuclear chart. However, since the $\beta$-decay proceeds via weak interaction, the process is slow and less favored and energy involved (released) is also less compared to spontaneous fission and $\alpha$ decay which proceed quickly via strong interaction making these processes more probable. For heaviest nuclei, the mutual repulsion of electric charge is higher than surface energy of the nucleus arising from the short range nuclear forces. On the basis of liquid drop model, Bohr and Wheeler \cite{bo39} described the mechanism of nuclear fission and established a limit for $Z^2/A\sim 48$ for spontaneous fission. Beyond this limit nuclei are unstable against spontaneous fission. According to this fact, no nucleus beyond $Z\sim 100$ can exist due to very small fission barrier. Both theoretical and experimental investigations on superheavy nuclei (SHN) support the fact that a bound superheavy nucleus can be formed only due to shell effects.\\

A simple semi-empirical formula on spontaneous fission half-lives for even-even, odd A and odd-odd nuclei in the ground state was proposed by  W.J. Swiatecki in 1955 \cite{sw55}. By including the deviation of experimental ground state masses from a smooth reference surface based on liquid drop model, Swiatecki successfully reproduced the experimental data with his semi-empirical formula \cite{sw55}. Microscopic calculation of spontaneous fission half-lives is very difficult due to both the complexity of the fission process and the uncertainty of the height and shape of the fission barrier \cite{mo89}. Ren and Xu \cite{xu05,ren05} generalized the formulae of spontaneous fission half-lives of even-even nuclei in their ground state to both, the case of odd nuclei and the case of fission isomers \cite{ren05}. The spontaneous fission half-lives of odd-A nuclei and of odd-odd nuclei in the ground state were calculated by using the generalized form of the Swiatecki's formula and by a new formula where the blocking effect of unpaired nucleon on the half-lives were taken into account. By introducing a blocking factor or a generalized seniority in the formulae of the half-lives of even-even nuclei, the experimental fission half-lives of odd-A nuclei and of odd-odd nuclei with Z=90 to Z=108 were reasonably reproduced with the same parameters used in ground state of even-even nuclei. 

In the present work, spontaneous fission half-lives for both neutron deficient and neutron rich isotopes of elements Z=102-120 have been calculated using the following formula \cite{ren05}:       
\begin{eqnarray}
log_{10}(T_{1/2}/yr) = 21.08 + c_1\frac{(Z - 90-v)}{A} \nonumber \\
+ c_2\frac{(Z -90-v)^2}{A} + c_3\frac{(Z -90-v)^3}{A} \nonumber \\
+ c_4\frac{(Z -90- v)}{A}(N -Z -52)^2 
\end{eqnarray}
where $c_1 =-548.825021$, $c_2 =-5.359139$, $c_3 = 0.767379$, $c_4 =-4.282220$,  $v$ = 0 for the spontaneous fission of even-even nuclei and $v$ = 2 for odd A and odd-odd nuclei. The seniority number $v$ was introduced for taking the blocking effect of unpaired nucleon on the transfer of many nucleon-pairs during the fission process. A variation of SF half lives calculated by this formula (eqn.5) with increasing neutron number for different elements from Z=102 to Z=120 are shown in Fig.4 and Fig.5 for comparison with the $\alpha$ decay and SF half lives predicted by our calculation and by Smolanczuk et al.\cite{sm95,sm97} respectively. Moreover, in this work, spontaneous fission half-lives calculated in a dynamical approach using macroscopic-microscopic method (MMM) of Ref.\cite{sm95,sm97} are also used (Figs.2-5) to find out the ``island" where the predicted SHE could survive the fission.\\   

\section{Results and Discussion}

In this paper, our main aim is to check whether the predicted SHN do really survive against the spontaneous fission (SF). Gamow-Teller $\beta$-decay half lives ($T_\beta$ in figs.2-3) obtained from a microscopic quasi-particle random phase approximation with single particle levels by Moller, Nix, Kratz \cite{MNK97} has also been used in this work to check the possibility of $\beta$-decay of SHN. In earlier works \cite{adndt07,jpsj,prc06,cs07,prc07}, we were able to well reproduce the available experimentally measured $\alpha$ decay half lives following the semi-classical quantum tunneling method with double folded density dependent effective M3Y interaction. In Ref. \cite{cs07}, $T_\alpha$ of about 314 nuclei were predicted using $Q_\alpha$ values from MMM ($Q_M$) \cite{mu01,mu103,mu203} and modified liquid droplet model of Myers-Swiatecki \cite{ms} to show the necessity of more accurate mass formula in determining the $Q_\alpha$-values with a good accuracy at least correct upto 10KeV for heaviest nuclei. But the spontaneous fission survivability of those nuclei was not checked in our previous work. For this reason, although some SHN (Z=102-120) are long-lived against $\alpha$-decay~\cite{prc06,cs07}, they may not live long if they have shorter SF half lives.
For example, at N=162 the magnitudes of $T_\alpha$ using $Q_M$ for the elements No (Z=102), Rf (Z=104), Sg (Z=106), Hs (Z=108) are $3.06\times 10^9s, ~4.10\times 10^6s, ~1.16\times 10^4s, ~28s$ respectively \cite{cs07}. If SF half lives ($T_{SF}$) of such nuclei are about the same order or relatively longer than their respective $T_\alpha$, only then a significant fraction of nuclei can survive fission and decay by $\alpha$ emission. In such cases $\alpha$-decay chain followed by SF of one of the product nuclei may be observed. Therefore, accurate estimation of $T_{SF}$ of SHN is essential to know whether predicted long-lived SHN against $\alpha$ emission do really exist against SF.\\

\begin{table*}
\caption{Comparisons between observed and theoretical (this work) $\alpha$-decay half lives using measured $Q_\alpha$. 
The experimental $\alpha$-decay half lives ($T_{1/2}^{EXP}$) are taken from Ref.\cite{gup05} except the values with single (*) and double (**) asterisk symbols which are taken from Ref. \cite{ho07} and Ref. \cite{dv06} respectively.}
\begin{ruledtabular}
\begin{tabular}{ccccccc}
Parent&Exptl Q-value&Half-lives (Exp)&This work($\ell=0$)&This work($\ell\neq 0$)\\
\hline 
$^AZ$&$Q_{ex}~(MeV)$ &$T_{1/2}^{EXP}$&$T_{1/2}^{DDM3Y}[Q_{ex}]$&$T_{1/2}^{DDM3Y}[Q_{ex}]$&$\ell$\\
\noalign{\smallskip}\hline\noalign{\smallskip}\\
$^{283}112$&$	9.704 \pm 0.015	$&$	^{*}6.9^{+6.9}_{-2.3}~s 	$&$	4.67^{+0.49}_{-0.44}~s	$&&&\\\\
$^{277}112$&$	11.594 \pm 0.055	$&$	0.69^{+0.69}_{-0.24}~ms $&$93^{+29}_{-23}~\mu s$& $0.59^{+0.19}_{-0.14}~ms$ &$4$\\\\
$^{272}111$&$	11.150 \pm 0.035	$&$	3.8^{+1.4}_{-0.8}~ms$ &$1.31^{+0.27}_{-0.22}~ms$ \\\\

$^{273}110$&$	11.37 \pm 0.05	$&$	0.17^{+0.17}_{-0.06}~ms  $&
$75^{+21}_{-17}~\mu s$ & $0.13^{+0.04}_{-0.03}~ms$ &$2	$\\\\
$^{271}110$&$	10.899 \pm 0.020	$&$	1.63^{+0.44}_{-0.29}~ms	$&$930^{+110}_{-90}~\mu s$ & $1.15^{+0.14}_{-0.12}~ms$ &$1	$\\\\

$^{270}110$&$	11.20 \pm 0.05  	$&$	0.10^{+0.14}_{-0.04}~ms  	$&$0.083^{+0.024}_{-0.019}~ms$ & $0.10^{+0.03}_{-0.02}~ms$ &$1	$\\\\

$^{269}110$&$	11.58 \pm 0.07	$&$	179^{+245}_{-66}~\mu s $&$28^{+13}_{-8}~\mu s$ 
& $88^{+36}_{-26}~\mu s$ &$3	$\\\\
$^{267}110$&$	12.28 \pm 0.11	$&$	2.8^{+13.3}_{-1.2}~\mu s$&$1.1^{+0.8}_{-0.4}~\mu s   	$\\\\
$^{268}109$&$	10.486 \pm 0.035$&$	21^{+8}_{-5}~ms           	$&$12.3^{+2.8}_{-2.2} ~ms	$\\\\

$^{266}109$&$	10.996 \pm 0.025$&$	1.7^{+1.8}_{-1.6}~ms	$&$750^{+110}_{-90}~\mu s$ & $1.3^{+0.2}_{-0.1}~ms$ &$2	$\\\\
$^{270}108$&$	9.30^{+0.07}_{-0.03}	$&$	3.6^{+0.8}_{-1.4}~s$&$	1.36^{+0.30}_{-0.51}~s$\\\\
$^{**270}108$&$	9.02 \pm 0.03                 $&$	^{**}22~s          	$&$	9.53^{+2.24}_{-1.86}~s   	$\\\\
$^{269}108$&$	9.315 \pm 0.022	$&$	9.7^{+9.7}_{-3.3}~s     	$&$	3.19^{+0.52}_{-0.43}~s  	$\\\\
$^{267}108$&$	9.978 \pm 0.020	$&$	58^{+23}_{-14}~ms 	$&$	45.5^{+5.8}_{-5.2}~ms	$\\\\
$^{266}108$&$	10.336 \pm 0.020	$&$	2.3^{+1.3}_{-0.6}~ms  	$&$	2.24^{+0.27}_{-0.25}~ms	$\\\\
$^{267}107$&$	8.96 \pm 0.30	$&$	17^{+14}_{-6}~s$&$	12^{+93}_{-11}~s$\\\\
$^{266}106$&$	8.88 \pm 0.03  	$&$	62^{+166}_{-44}~s    	$&$	4.89^{+1.20}_{-0.93}~s$& $16^{+4}_{-3}~s$ &$3	$\\
\end{tabular}
\end{ruledtabular}
\end{table*}

Gupta and Burrows \cite{gup05} summarised the measured ground state values of spontaneous fission and $\alpha$ decay half lives with $Q_\alpha$ for heaviest nuclei having mass number A=266-294. These values were taken from experimental measurements carried out at GSI, Germany, and JINR, Dubna. In Table I, a comparison between experimentally measured and calculated half lives using DDM3Y effective nucleon-nucleon (NN) interaction in a WKB framework is shown for sixteen superheavy nuclei. Most of these nuclei (twelve out of sixteen) were not addressed in our earlier works \cite{adndt07,jpsj,prc06,cs07,prc07}. In Fig.1 the theoretical $Q_\alpha$-values from different mass formulae prescribed by MS \cite{ms}, KUTY\cite{kuty00}, Muntian et al. (MMM) \cite{mu01,mu103,mu203} are employed to calculate $\alpha$-decay half-lives of the same sixteen SHN presented in Table I. The measured $Q_\alpha$-values are used to calculate $\alpha$ decay half lives ($T_{1/2}^{DDM3Y}[Q_{ex}]$) in Table I. Half lives calculated in the present work agree reasonably well with experimental data for most of the superheavies. Nuclei for which spin-parities of parent and daughter nuclei are not known, zero angular momenta ($\ell=0$) transfers are used. Since $\ell=0$ gives minimum centrifugal barrier, probability of $\alpha$-tunneling increases and consequently half life decreases. But, for some nuclei like $^{277}112$, $^{273}110$, $^{269}110$, $^{266}106$ etc. angular momenta carried by $\alpha$ particles may not be zero. For these nuclei, calculated half lives are too small compared to measured values. In particular, for $^{277}112$ calculated value ($\sim 0.093~ms$) is less than the measured one ($0.69~ms$) by one order of magnitude. This discrepancy can be removed only if one can assume non-zero orbital angular momentum transfer in the $\alpha$ decay process. In our calculation, $\ell=4$ gives the better agreement of predicted half life for $^{277}112$ nucleus with measured value. It is therefore important to determine the proper spin-parity of parent and daughter nuclei to enable correct centrifugal barrier for $\alpha$ decay half life calculation. Although, higher $\ell$-values are shown for some nuclei for better agreement with the experimental result, but more experiments on the same nuclei (shown in table I) with higher statistics are needed for reconfirmation of the data and measured $\alpha$ decay energies since reaction cross-sections are of the order of pico-barns (pb). \\

    In this work, existing SF calculation with a microscopic approach \cite{sm95,sm97} have been used to find the region of long lived fission survived nuclei in the SHN region of the nuclear chart. In Fig.2 and Fig.3, comparisons between calculated $T_\alpha$, $T_\beta$ and $T_{SF}$ are shown only for even-Z elements with proton number Z=102-120 since $T_\alpha$ and $T_{SF}$ calculations of  Ref.\cite{sm95,sm97} are valid only for even-even nuclei. A few available observed data for both $T_\alpha$ and $T_{SF}$ are also shown for most of the elements in the plots of Figs 2-3. For Z=104, the highest value of $T_\alpha$ ($\sim 4.1\times 10^6 s$) according to calculation of this work using $Q_M$ appears around N=162 where much smaller value of $T_{SF}$ calculated in Ref. \cite{sm95} ($\sim 23~s$) makes this nucleus $^{266}104_{162}$ unstable against SF. Therefore, if synthesis of this nucleus is possible in the present-day setup, substantial fraction of $^{266}104_{162}$ would undergo SF within a few seconds. For Z=102, no such calculation on SF is available in Ref.\cite{sm95,sm97}. Since calculated $T_{SF}$ and $Q_M$ values both are based on MMM, $\alpha$ decay half life calculations using DDM3Y effective interaction with $Q_M$ ($T_\alpha$~M3Y~Q-M) is preferably chosen to compare with $T_{SF}$. In addition to that, calculation within the same framework using $Q_{KUTY}$ have also been presented in Figs 1-3.\\

In case of Sg (Z=106) isotopes, DDM3Y with  $Q_M$ predicts that the longest $T_\alpha$ ($\sim 1.16\times10^4 s\sim3.2~hr$) would be at N=162. It is comparable to $T_{SF}$ (3.5~hr) of Ref.\cite{sm95}.
Therefore, $\alpha$ decay channel is one of the dominant decay mode of this $^{268}Sg_{162}$ nucleus. If it is produced in the laboratory, it may be observed only for few hours ($ lifetime \sim 1.68~hr$) since both SF and $\alpha$ decay half lives are small. It is to be noted that in the present work, lifetime of some SHN (either $\beta$-stable or have large $T_{\beta}$)are predicted by considering SF and $\alpha$ decay half lives only. Incidentally, $^{264}No_{162}$, $^{266}Rf_{162}$, $^{268}Sg_{162}$, and $^{270}Hs_{162}$ are known to be either $\beta$-stable or have very large $T_\beta$ \cite{MNK97}.
$Q_\alpha$ values using KUTY mass formula are not always very much reliable since it does not reproduce all the observed Q values with a good accuracy. But, this mass formula can be used to locate the region of possible existence of long-lived SHN where Q values from other mass formula are not available. 
It may be pointed out that the use of $Q_{KUTY}$ in our calculation shows reasonable agreement for several nuclei in Fig. 1. \\

   As $Q_M$ values are not available for more neutron rich isotopes of Sg, using $Q_{KUTY}$ in present calculation predicts $T_\alpha \sim 3.71\times10^{15}~s \sim 1.18\times 10^8~yrs$ at N=184. The SF half life of $^{290}Sg_{184}$ is less ($T_{SF}\sim 4.07\times10^{13}~s \sim 1.3\times10^6~yrs$) than $T_\alpha$. Although $T_\beta$ of this nucleus is not specified but it is expected to be large in ref.\cite{MNK97}. Hence, if synthesized, $^{290}Sg_{184}$ is expected to have long enough lifetime ($\tau\sim 1.27\times 10^6~yrs$) to be observed in the laboratory. But it is still smaller than the age of earth $\sim 4.5\times 10^9$ yrs by three orders of magnitude. Calculated $T_\alpha$  using $Q_M$ $(28~s)$ for $^{270}Hs_{162}$ is less than   $T_{SF}$ ($\sim1.8~hr$) and therefore $\alpha$ decay chain of such nucleus is expected to be observed. This prediction from our calculation is in good agreement with recently observed \cite{dv06} doubly magic deformed nucleus $^{270}Hs_{162}$. Present calculation using experimental Q-value gives $T_\alpha \sim 9.53 s$ (Table I).\\
           
   From Figs 2-3 it is seen that the extra stability effect of neutron shell at N=162 almost disappears with the increase of atomic number and $T_\alpha$ becomes of the order of millisecond to microsecond for the elements having $Z\ge 110$. On the other hand, in more neutron rich side around N=184 of elements Z=110, 112, 114  theoretical $T_\alpha$ using $Q_{KUTY}$ in present calculation are of the order of $10^{10}s, 10^{8}s, 10^{6}s$ respectively which are much less than theoretical $T_{SF}$ of the order of $10^{12}s, 10^{13}s, 10^{13} s$ respectively. It must be noted that $^{296}112_{184}$, $^{298}114_{184}$ are $\beta$-stable whereas $^{294}110_{184}$ is predicted to have large $T_\beta$ in ref.\cite{MNK97} due to its very small positive $Q_\beta$-value. $\beta$-stable nuclei and those with very large $T_\beta$ are not shown in figs 2-3. \\

  Beyond Z=114 peak-value of $T_\alpha$ plot (Fig.3) around N=184 suddenly reduces showing a possible signature of spherical proton shell at Z=114. According to the present calculation, $^{298}114_{184}$ the so-called doubly magic spherical superheavy nucleus predicted by nuclear structure theory in the mid of 1960, has $T_\alpha$ values of the order of $10^6s$ and $5\times 10^2s$ using $Q_{KUTY}$ and $Q_{M}$ respectively which are much less than $T_{SF}\sim10^{13}s$.\\

   In Figures 4-5, three curves describing spontaneous fission and $\alpha$-decay half lives are shown in each graph excepting Z=102 for which calculation of spontaneous fission half lives in Ref.\cite{sm95,sm97}, represented by Tsf~SM in figs.~4-5, are not available. For the isotopes of elements Z=106, 108 the values of Tsf~SM become of the order of one millisecond near N=170. However, the values of Tsf~SM are comparable to the calculated $\alpha$ decay half lives of this work ($T_\alpha$~M3Y~Q-M) around N=154 to 164 for Sg. In case of Hassium isotopes (Z=108, N=156-163), Tsf~SM $>$ $T_\alpha$~M3Y~Q-M reveals the fact that isotopes of Hs within this range may undergo $\alpha$ decay with a longest half life of the order of few seconds ($\sim$10-30 s). This is in good agreement with the recent experimental observation of $\alpha$ decay from the nucleus $^{270}Hs_{162}$. But spontaneous fission half lives calculated using Eqn.(5) \cite{xu05,ren05} fall rapidly around N=160-170 for elements having Z=102-112. In fig.5, SF half life values for neutron rich isotopes of elements having Z=114, 116, 118, 120 are less than $10^{-15}s$ near N=180. For Z=114 only, this calculation matches with microscopic results for some isotopes having N=160-170. This calculation contradicts the following two facts:
(i) Experimentally measured value of $T_\alpha$ for $^{283}112_{171}$ is $\sim 6.9~s$ (see Table I) whereas, SF half life for this nucleus calculated by using equation (5) is extremely low ($\sim 10^{-11}~s$) indicating immediate fissioning of the nucleus. (ii) Since $T_{SF}$ by Ren and Xu rapidly falls around N=160-170 (fig.4) and N=180 (fig.5) for Z=102-112 and Z=114-120 respectively, it could not explain the predicted extra shell stability of SHN at N=184 for heavier elements.\\

          From the present calculation of $T_\alpha$ using $Q_{KUTY}$ it seems that longer $T_\alpha$ might be observed for more neutron rich side ($Z\ge 116$, $N>190$) due to possible existence of neutron shell closure. On the contrary, from the trend of SF plots for Z=116, 118, 120 it appears that lowering of $T_{SF}$ might destroy the existence of such neutron rich ($N>190$) superheavy isotopes of elements Z=116, 118 and 120. Hence, the presence of long-lived SHN with neutron shell closure beyond N=184 may be ruled out. However, it is an important task to determine the SF and $\alpha$ decay half lives for $N>184$ to confirm whether there is any possibility of neutron shell closure beyond N=184 for heaviest elements ($Z\ge 116$).\\

   In two graphs of Fig. 6, using $Q_M$ values in this calculation, variation of $\alpha$ decay with neutron number for both odd-Z and even-Z elements having Z=102-120 are shown. Plots of both graphs (a) and (b) of Fig.6 clearly show peaks of $T_\alpha$-values around N=162 and N=184 for all elements with Z=102-120, which possibly indicates the neutron shell closure at N=162 and 184. This is in good agreement with the present-day knowledge of microscopic theory.    
      
\section{Summary and conclusion}
\noindent
The natural existence of superheavy nuclei is limited primarily by spontaneous $\alpha$ decay and spontaneous fission processes. A SHN in spite of having longer $\alpha$ decay half lives may undergo immediate spontaneous fission if the latter has a low half life. On the other hand, SF stable SHN may have shorter ($\sim$ 1$\mu$s or less) $\alpha$ decay half lives. In both the cases, such SHN may not be observed even if they are synthesized in the present day laboratory setup. The main aim of this work is to find out the fission-survived long lived SHN. In fact, if SHN have high degree of stability against both $\alpha$ decay and SF, we would be able to observe them if produced in the laboratory provided those SHN are not far away from $\beta$-stability line. We have calculated the $\alpha$ decay half lives of SHN in quantum tunneling method with microscopic NN potential using Q-values from different mass formulae and compared them with the $\beta$-decay and SF half lives to find the long lived SHN.\\

The highlights of observations made in this work are summarised as follows:

       (i) Among all three mass formulae, $Q_\alpha$-values used from MMM model ($Q_M$) \cite{mu01,mu103,mu203} in the present method, reproduces the observed data reasonably well (see fig.1), but non-availability of $Q_\alpha$-values in more neutron rich side limits its usage. Therefore, for higher Z region, the mass formula of KUTY, which extends up to Z=130, has been considered.

       (ii) Although $^{266}Rf_{162}$ nucleus has relatively longer $T_\alpha$ half life ($\sim 4.1\times 10^6 s \sim $47.5 days using $Q_M$) but it is unstable against SF with Tsf~SM$\sim 23s$ 
only (Fig. 2). 

       (iii) The mass formula of KUTY predicts $\alpha$ decay life time of the $^{290}Sg_{184}$ nucleus to $\sim 10^{8}$ yrs whereas  Tsf~SM$\sim 10^6~yrs$ makes the lifetime ($\sim 10^6 yrs$) of this nucleus very long but still smaller than the age of the earth ($\sim 4.5\times 10^9$yrs) by three orders of magnitude. This nucleus is either $\beta$-stable or might have very large $T_\beta$ according to the calculations of ref.\cite{MNK97}.

       (iv) The larger deviations between calculated and experimentally measured $\alpha$ decay half lives are observed in case of only few nuclei such as $^{277}112$ which may be due to higher minimum orbital angular momenta carried away by $\alpha$ particles for spin-parity conservation. Inaccuracy in the measurement of $T_\alpha$ of $^{277}112$ nucleus due to very low count rate also may not be ruled out.

       (v) Using the formulation based on liquid drop model in Ref.\cite{xu05,ren05} SF half lives calculation have also been done for higher Z elements in this work.  Results shown in plots of figs.4-5 do not match with SF half lives calculation in a microscopic approach of Ref.\cite{sm95,sm97} for both neutron rich and neutron deficient isotopes of heaviest elements with Z=104-120. The half life value using this phenomenological prescription also contradicts the observed $\alpha$-decay from $^{283}112_{171}$ nucleus with measured $T_\alpha \sim 6.9 s$.

       (vi) It is evident from the present $T_\alpha$ calculations that the effect of the deformed neutron shell closure at N=162 will be insignificant beyond Z=108 as $T_\alpha$ goes on decreasing with increasing atomic number. For N=162 isotopes of elements having Z=110, 112, values of $T_{SF}$ are $9.8m$, $0.63s$ respectively and $T_\alpha$ using $Q_M$ are of the order of milliseconds ($\sim 1ms$) and microseconds ($\sim 93\mu s$) (see plots of  Fig.3) respectively i.e. the values of $T_\alpha$ go on decreasing more rapidly than the corresponding SF half lives with increasing atomic number.   

       (vii) Calculated $T_\alpha$ using $Q_{KUTY}$ predicts almost $\beta$-stable long lived SHN around $^{294}110_{184}$, $^{296}112_{184}$, $^{298}114_{184}$ with $T_\alpha$ of the order of $\sim 311yrs, \sim 3.10yrs, \sim 17 days$ respectively which are much less than their $T_{SF}$ ($\sim 4.48 \times 10^4 yrs, \sim 3.09\times 10^5 yrs, \sim 4.38\times 10^5 yrs$ respectively) values. Hence the dominant decay mode of the above nuclei and their immediate neighbours is expected to have $\alpha$ decay mode. $T_\alpha$ value of $^{293}110_{183}$ is about 352 years which is slightly greater than that for $^{294}110_{184}$ nucleus. The SF half life of this nucleus is not found in Ref.\cite{sm95,sm97}. For $^{292}108_{184}$ nucleus since $T_\alpha$ ($\sim 9.6\times 10^4 yrs $ using $Q_{KUTY}$) value is comparable to its $T_{SF}$ ($\sim 3.2\times 10^4 yrs$), this nucleus is one of the possible members of stability island. The exact values of $T_\beta$ of $^{293}110_{183}$ and $^{292}108_{184}$ nuclei are not shown in ref.\cite{MNK97} but predicted to be large.

       (viii) Using $Q_{KUTY}$ values in the present calculation shows longer $T_\alpha$-values for neutron rich ($N>190$) isotopes of Z=116, 118 and 120 indicating a possible neutron shell closure next to N=184 might occur. On the contrary, calculated SF half lives (Fig.3) of Ref.\cite{sm95,sm97} show a trend of lowering of $T_{SF}$ ($<1ms~ to~1\mu s$) for neutron rich isotopes of those elements which indicates the higher probability of SF of such SHN in this region. However, more accurate determination of fission barrier and their corresponding half lives are essential to predict long lived SHN in the region of very high atomic number.   

       (ix) It may be pointed out that the calculation of $T_\alpha$ is very sensitive to $Q_\alpha$-values, and none of the mass formula used here cover the entire mass range with extreme accuracy. Therefore, a better mass estimate covering the wide range of superheavy masses with a good accuracy is necessary.\\

In summary, we find that the possibility of existence of SHN above Z= 114 with considerable life time is very low. Although Z=120, 124, 126 with N=184 might form spherical-doubly-magic nuclei and survive fission \cite{st06}, they would undergo $\alpha$ decay within microseconds. A small ``island/peninsula" might survive fission and $\beta$-decay but undergo $\alpha$ decay in the region Z=106-108, N $\sim$ 160-164. Interestingly, in this region the $\beta$-stable SHN Z=106, N=162 has the highest $\alpha$ decay half life $\sim 3.2 hrs$ (Fig. 2, using $Q_M$) that is
much greater than the recently discovered deformed-doubly-magic SHN
$^{270}$Hs (measured $T_{\alpha}\sim$ 22 secs). Thus a search for
this long-lived SHN $^{268}Sg_{162}$ can be pursued. Similarly, the
nucleus with Z=110, N=183 appears to be near the center of a possible
``magic island" (Z=104-116, N $\sim$ 176-186) with $\alpha$ decay half
life $\sim 352 yrs$ (Fig. 3, using $Q_{KUTY}$) which is greater than
that of the doubly-magic SHN Z=114, N=184 ($T_\alpha \sim 17$days).
Since the SHN $^{290}Sg_{184}$ has $T_\alpha$ and $T_{SF}$ values $\sim
10^8$ yrs (Fig. 2, using $Q_{KUTY}$), and $\sim 10^6$ yrs
respectively, it might have longer life time in comparison to other superheavies. However, for both $^{293}Ds_{183}$ and $^{290}Sg_{184}$ nuclei, $\beta$-decay might be another possible decay mode with large $T_\beta$ values. Only future experiments can confirm this. Finally, the experimental investigations to detect the $\alpha$-cascade can be pursued on $^{294}110_{184}$, $^{293}110_{183}$, $^{296}112_{184}$ and $^{298}114_{184}$ nuclei which are expected to decay predominantly through $\alpha$ particle emission.

\pagebreak

\begin{figure*}[h]
\eject\centerline{\epsfig{file=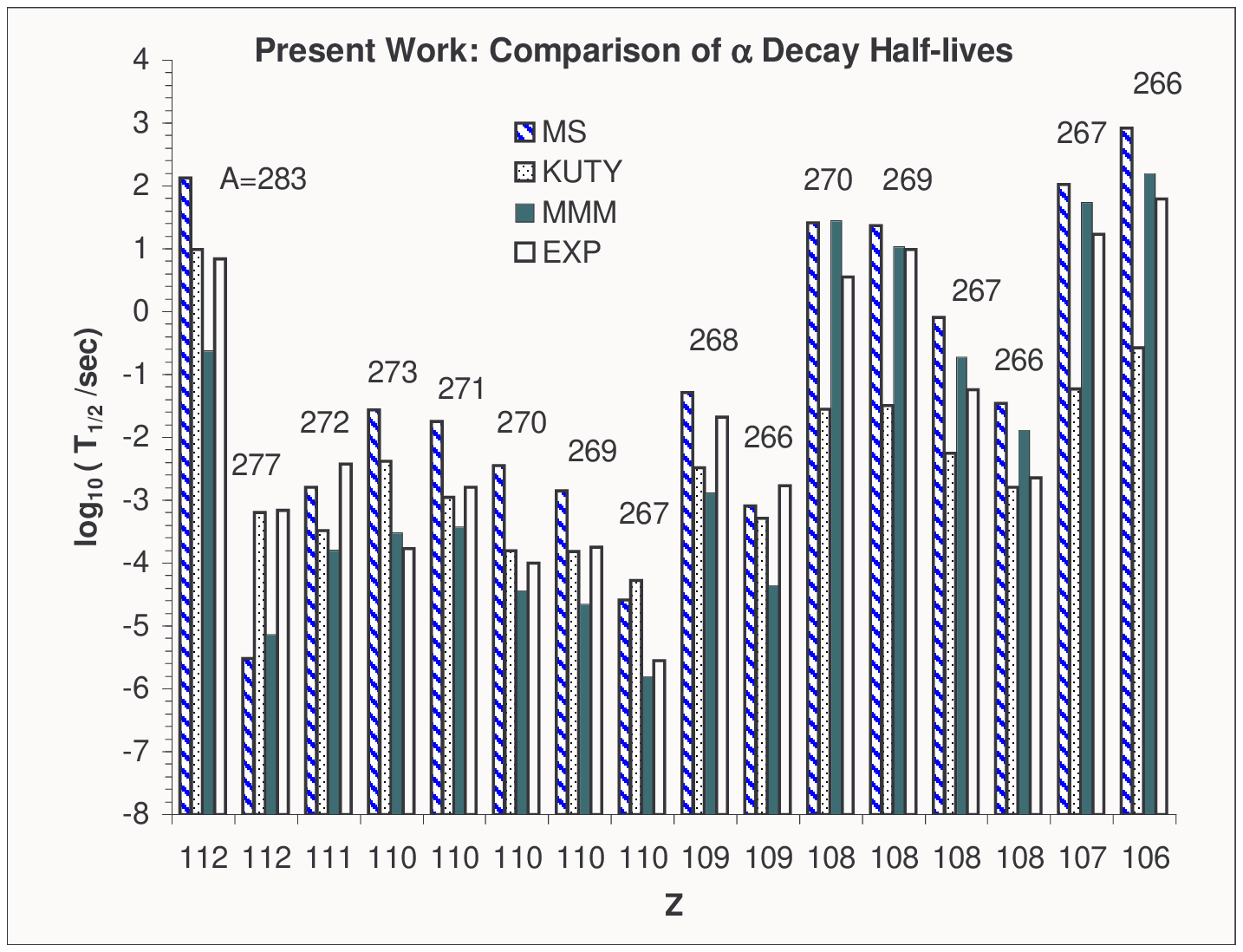,height=12cm,width=16cm}}
\caption 
{``(Color online)" Plots of $\alpha$ decay half life [$log_{10}(T_{1/2}/sec)$] in logarithmic scale versus proton number Z for different mass number A (indicated  on top of each coloumn) using zero angular momentum transfer ($\ell=0$). (a) bar coded columns are theoretical half lives ($T_\alpha$ M3Y Q-MS) in WKB frame work with DDM3Y interaction and [$Q^{MS}_{th}$] from Myers-Swiatecki mass formula, (b) columns filled with dots 
($T_\alpha$ M3Y Q-K) are in the same framework but with [$Q^{KUTY}_{th}$] from Koura-Tachibana-Uno-Yamada mass estimates, (c) solid columns are theoretical half lives 
($T_\alpha$ M3Y Q-M) in WKB frame work with DDM3Y interaction and [$Q^{M}_{th}$] from Muntian-Patyk-Hofmann-Sobiczewski mass formula, (d) hollow columns are experimental $\alpha$ decay half lives ($T_\alpha$ Expt). Experimental errors are given in Table I.}
\label{fig1}
\end{figure*}

\begin{figure*}
\eject\centerline{\epsfig{file=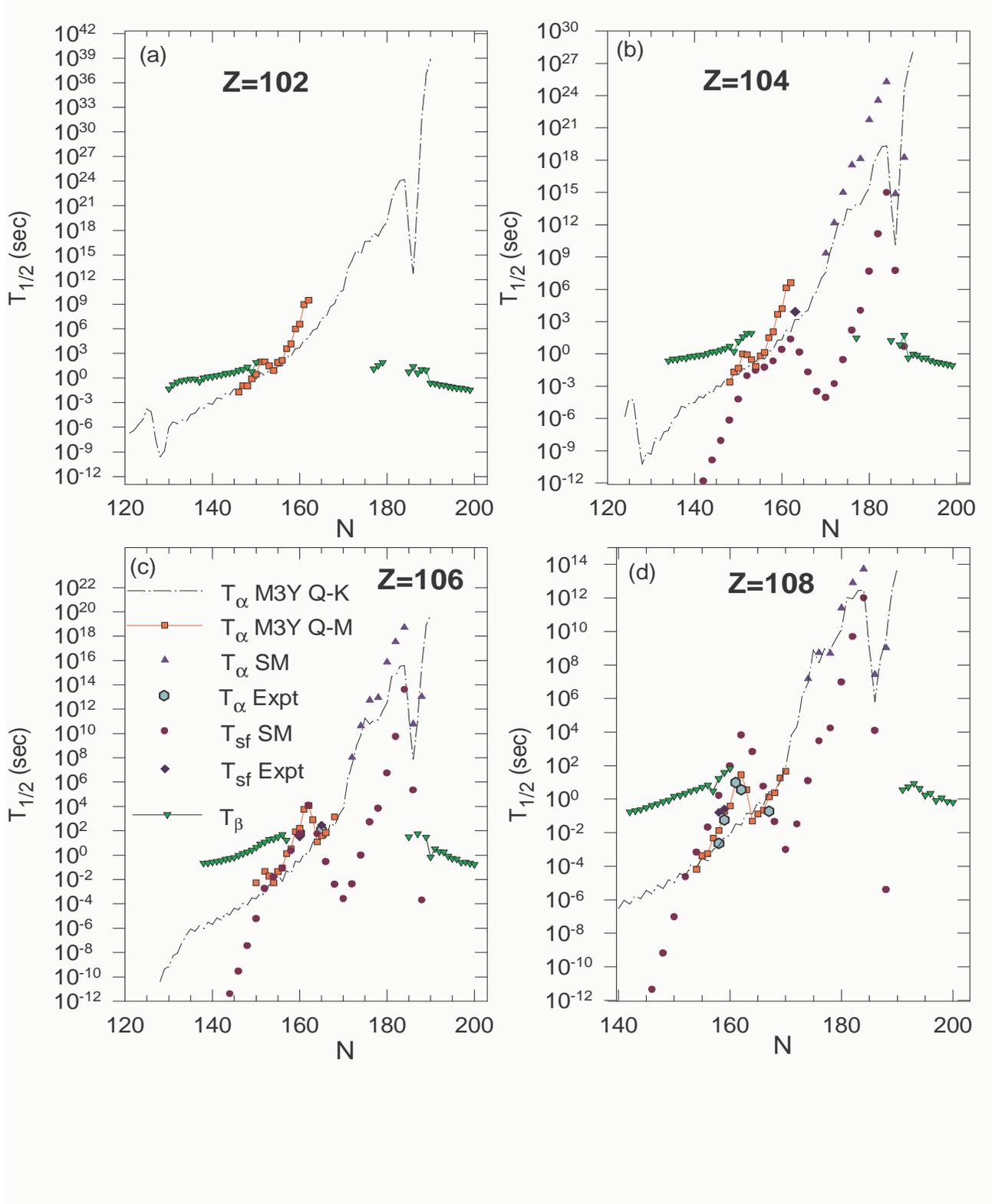,height=21cm,width=17cm}}
\caption{``(Color online)" Variation of $\alpha$ decay and fission half-lives with neutron number for elements (a) Z=102, (b) Z=104, (c) Z=106, (d) Z=108 are shown. For all plots the following symbols are used: Dash-dotted line ($T_\alpha$~M3Y~Q-K) and continuous line with square symbols ($T_\alpha$~M3Y~Q-M) represent $\alpha$ decay half-lives calculation using Q-values from KUTY (Q-K) and Muntian et al. (Q-M) respectively in this work. Triangle symbol ($T_\alpha$~SM) represents $\alpha$-decay half-lives predicted within a microscopic framework \cite{sm95,sm97}. Hexagon symbol ($T_\alpha$~Expt) represents measured $\alpha$ decay half-lives. Solid circle (Tsf~SM) represents fission half-lives predicted by microscopic calculation. Diamond symbol (Tsf expt) represents measured fission half-lives for some nuclei. Line-inverted triangle ($T_\beta$) shows $\beta$ decay half-lives predicted in ref. \cite{MNK97}. }
\label{fig2}
\end{figure*}

\begin{figure*}
\eject\centerline{\epsfig{file=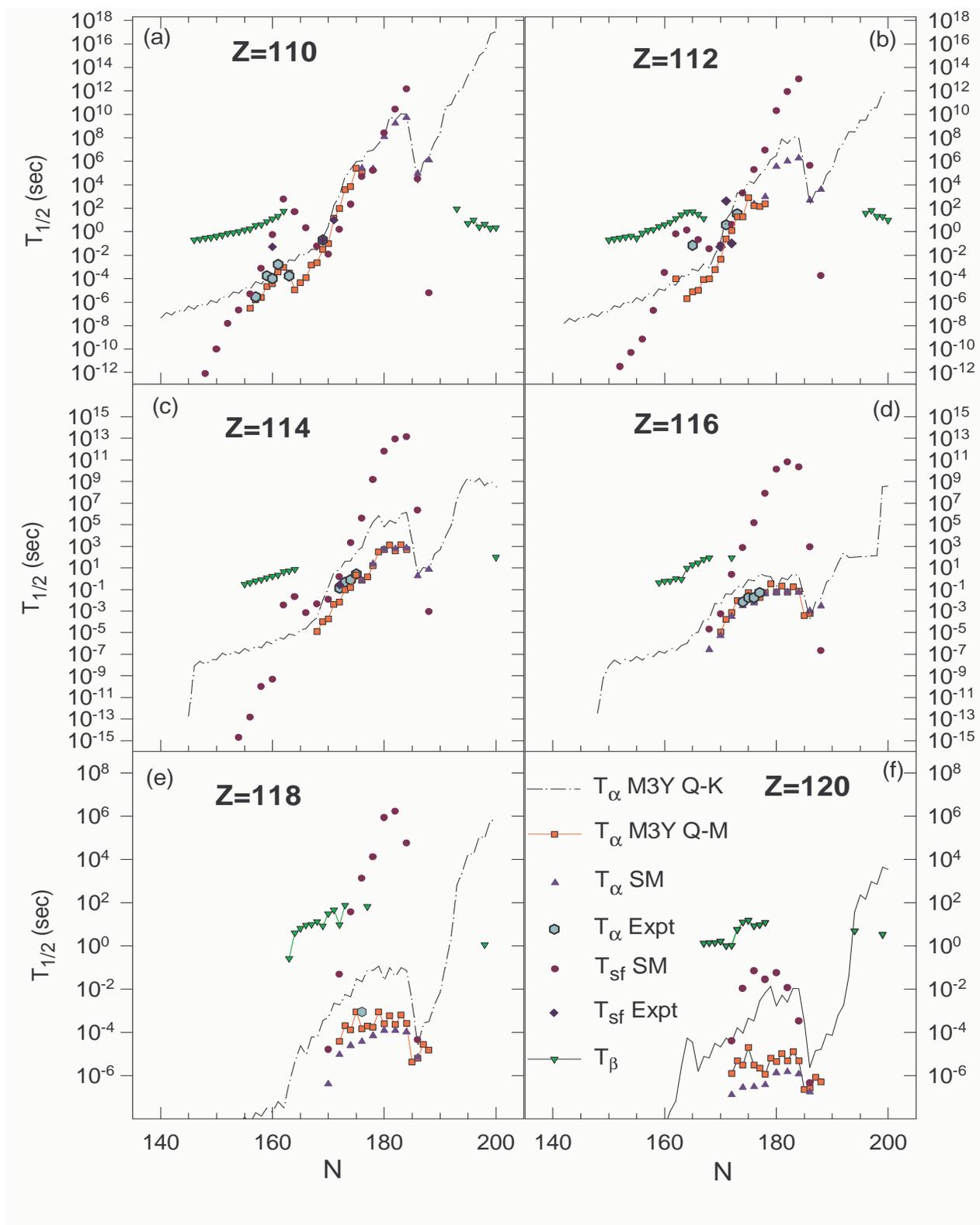,height=21cm,width=17cm}}
\caption{``(Color online)" Same as Fig.2  for elements (a) Z=110, (b) Z=112, (c) Z=114, (d) Z=116, 
(e) Z=118, (f) Z=120}
\label{fig3}
\end{figure*}

\begin{figure*}
\eject\centerline{\epsfig{file=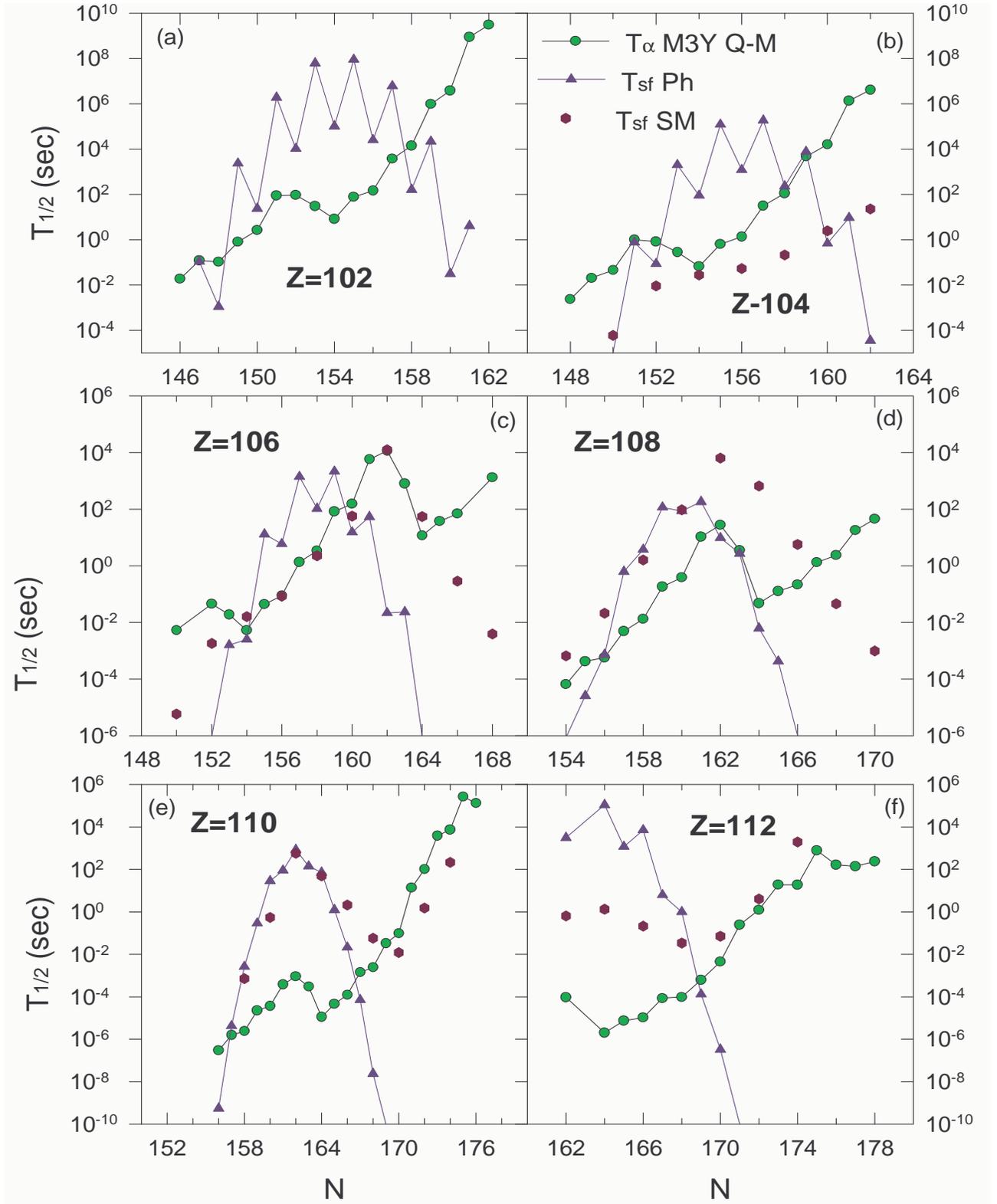,height=21cm,width=17cm}}
\caption{``(Color online)" Plots for variation of phenomenological \cite{xu05,ren05} and microscopically \cite{sm95,sm97} calculated fission half lives with neutron numbers for (a) Z=102, (b) Z=104, (c) Z=106, (d) Z=108, (e) Z=110, (f) Z=112. The corresponding $\alpha$ decay half-lives from the present calculation are also shown for comparison. Continuous line with solid circle ($T_\alpha$~M3Y~Q-M) represents $\alpha$ decay half-lives predicted by this work using Q value from Muntian et al.\cite{mu01,mu103,mu203}. Continuous line with solid triangle (Tsf~Ph) represents spontaneous fission half-lives predicted by phenomenological calculation. Solid dark circle (Tsf~SM) shows the spontaneous fission half-lives predicted by microscopic calculation.  }
\label{fig4}
\end{figure*}

\begin{figure*}
\eject\centerline{\epsfig{file=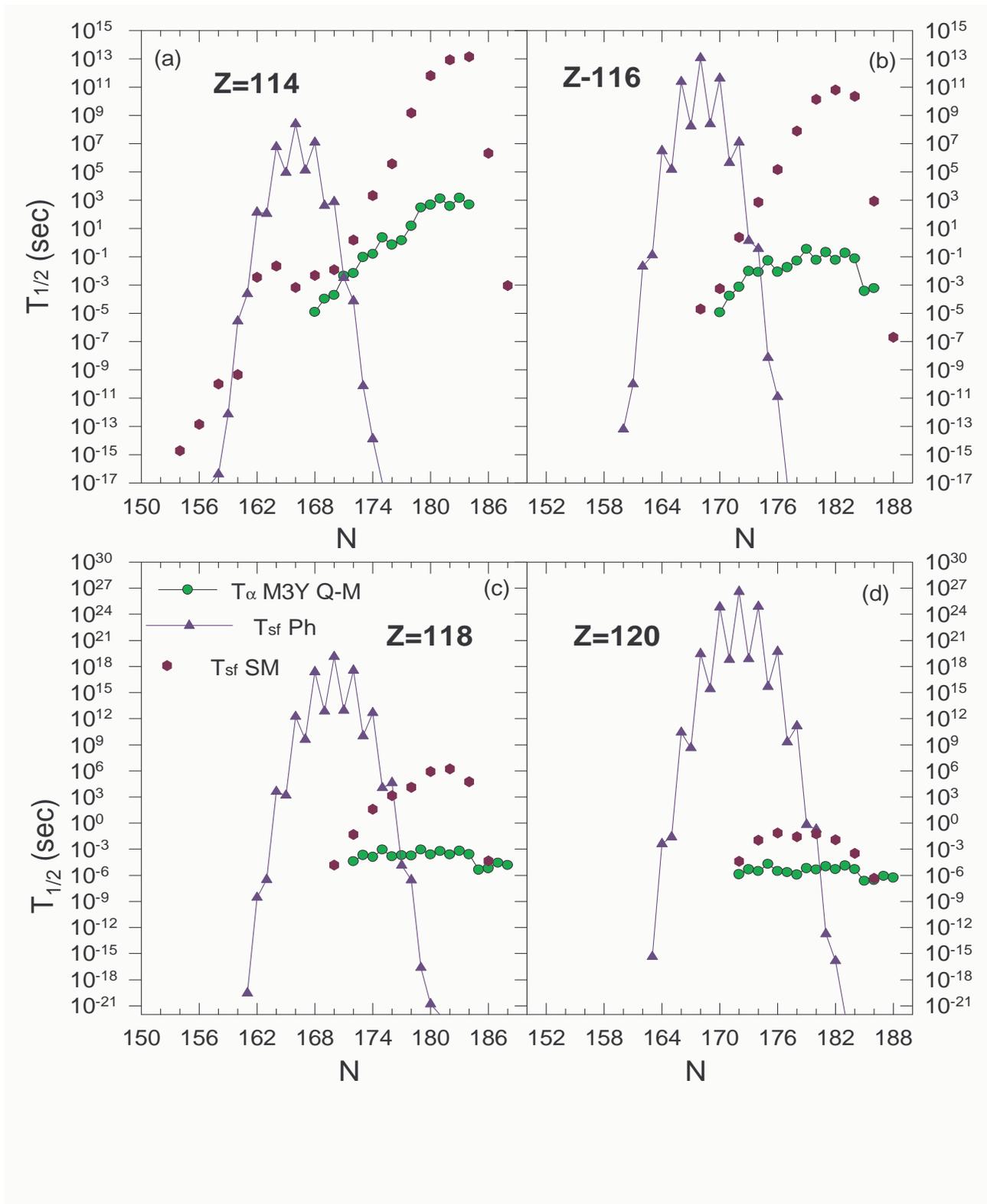,height=21cm,width=17cm}}
\caption{``(Color online)" Same as fig.4 for (a) Z=114, (b) Z=116, (c) Z=118, (d) Z=120 }
\label{fig5}
\end{figure*}

\begin{figure*}
\eject\centerline{\epsfig{file=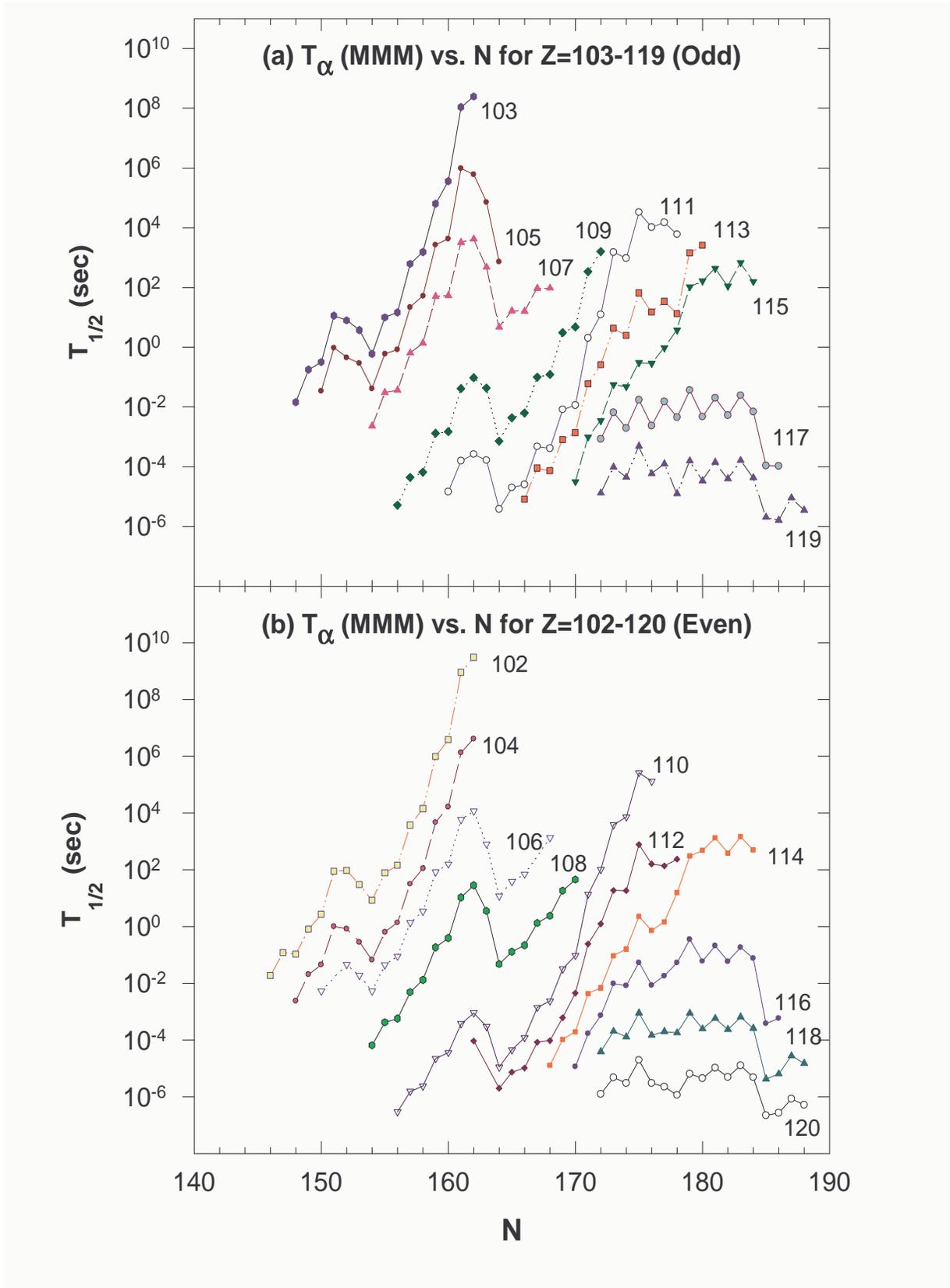,height=22cm,width=16cm}}
\caption 
{``(Color online)"  Variation of $\alpha$ decay half-lives ($T_\alpha$) predicted by present calculation with neutron number for (a) odd-Z and (b) even-Z elements from Z=102-120. Each graphs show the abrupt reduction of $T_\alpha$ around N=162 as the possible signature of shell effect. Similar reductions of $T_\alpha$ around N=184 are also shown by the elements for which $Q_\alpha$ values are available from MMM calculation.}
\label{fig6}
\end{figure*}

\end{document}